\documentclass[final,journal,letterpaper,twoside,onecolumn]{IEEEtran}

\usepackage[utf8]{inputenc} 
\usepackage[T1]{fontenc}
\usepackage{url}              

\usepackage[cmex10]{amsmath}  
\usepackage{mleftright}       
\mleftright                   

\usepackage{amsmath,amssymb,amsfonts, balance, mathtools,graphicx,textcomp,xcolor,amsthm, xpatch, tikz, forest, epstopdf, pdftexcmds, floatrow, subfiles, csquotes}

\usepackage[inline]{enumitem}

\usetikzlibrary{positioning}
\usepackage{booktabs}         

\usepackage[hidelinks]{hyperref}

\usepackage{orcidlink} 

\newcounter{cases}
\newcounter{subcases}[cases]

\makeatletter
\xpatchcmd{\proof}{\topsep6\p@\@plus6\p@\relax}{}{}{}
\makeatother

\newtheorem{constr}{Construction}

\newtheorem{theorem}{Theorem}
\newtheorem{lemma}{Lemma}
\newtheorem{corollary}{Corollary}

\newtheorem{defn}{Definition}
\newtheorem{env_example}{Example}

\newtheorem*{lemma*}{Lemma}
\newtheorem*{theorem*}{Theorem}

\DeclarePairedDelimiter{\ceil}{\lceil}{\rceil}
\DeclarePairedDelimiter{\floor}{\lfloor}{\rfloor}

\DeclarePairedDelimiter\abs{\lvert}{\rvert}
\DeclarePairedDelimiter\parenv{\lparen}{\rparen}

\DeclarePairedDelimiter\bracenv{\lbrace}{\rbrace}
\DeclarePairedDelimiter\floorenv{\lfloor}{\rfloor}
\DeclarePairedDelimiter\ceilenv{\lceil}{\rceil}
\DeclarePairedDelimiterX\mathset[2]{\lbrace}{\rbrace}{#1 : #2}

\newcommand{\cQ}{\mathcal{Q}}
\newcommand{\cR}{\mathcal{R}}

\newcommand{\bft}{{\boldsymbol t}}
\newcommand{\bfu}{{\boldsymbol u}}
\newcommand{\bfv}{{\boldsymbol v}}
\newcommand{\bfw}{{\boldsymbol w}}
\newcommand{\bfx}{{\boldsymbol x}}
\newcommand{\bfy}{{\boldsymbol y}}
\newcommand{\bfz}{{\boldsymbol z}}

\newcommand{\bfH}{{\mathbf H}}

\DeclareMathOperator{\RLL}{RLL}

\newcommand{\tr}[2][]{\mathcal{R}_{#1}(#2)}
\newcommand{\tri}[3][]{\tr[#1]{\boldsymbol{#2}}_{#3}}
\newcommand{\dtr}[3][]{\Delta_{#1}^{#3}({#2})}

\newcommand{\trw}[2][]{\mathcal{R}_{#1}^{\pi}(#2)}

\newcommand{\wread}{read }
\newcommand{\cnull}{\phi }

\IEEEoverridecommandlockouts

\begin{document}
\title{Error-Correcting Codes for Nanopore Sequencing}

\author{%
  Anisha~Banerjee\,\orcidlink{0000-0003-2285-1482}\,%
		,~\IEEEmembership{Student~Member,~IEEE}, 
  Yonatan~Yehezkeally\,\orcidlink{0000-0003-1652-9761}\,%
		,~\IEEEmembership{Member,~IEEE},
  Antonia~Wachter-Zeh\,\orcidlink{0000-0002-5174-1947}\,%
		,~\IEEEmembership{Senior~Member,~IEEE},
  and Eitan~Yaakobi\,\orcidlink{0000-0002-9851-5234}\,%
		,~\IEEEmembership{Senior~Member,~IEEE}
  \thanks{%
  Manuscript received 2~October~2023%
  . 
  This material is based upon work supported by the National Science Foundation under Grant No. CCF 2212437. This work has also received funding from the European Research Council (ERC) under the European Union’s Horizon 2020 research and innovation programme (Grant agreement No. 801434). It was also funded by the European Union (ERC, DNAStorage, 865630). Views and opinions expressed are however those of the author(s) only and do not necessarily reflect those of the European Union or the European Research Council Executive Agency. Neither the European Union nor the granting authority can be held responsible for them. 
  The work of Yonatan~Yehezkeally was supported by the Alexander von Humboldt Foundation under a Carl Friedrich von Siemens Post-Doctoral Research Fellowship. 
  An earlier version of this paper was presented in part at the 2023 {IEEE} International Symposium on Information Theory ({ISIT}) [\textsc{DOI:\; 10.1109/ISIT54713.2023.10206710}]. 
  \emph{(Corresponding author: Anisha~Banerjee.)}}
  \thanks{%
  Anisha~Banerjee, Yonatan~Yehezkeally, and Antonia~Wachter-Zeh are with the Institute for Communications Engineering, School of Computation, Information and Technology, Technical University of Munich, 80333 Munich, Germany 
  (e-mail: \texttt{anisha.banerjee@tum.de}; \texttt{yonatan.yehezkeally@tum.de}; \texttt{antonia.wachter-zeh@tum.de}). 
  Eitan~Yaakobi is with the Department of Computer Science, Technion---Israel Institute of Technology, Haifa 3200003, Israel 
  (e-mail: \texttt{yaakobi@cs.technion.ac.il}).}
}

\maketitle



\begin{abstract}

Nanopore sequencing, superior to other sequencing technologies for DNA storage in multiple aspects, has recently attracted considerable attention. Its high error rates, however, demand thorough research on practical and efficient coding schemes to enable accurate recovery of stored data. To this end, we consider a simplified model of a nanopore sequencer inspired by Mao \emph{et al.}, incorporating intersymbol interference and measurement noise. Essentially, our channel model passes a sliding window of length \(\ell\) over a \(q\)-ary input sequence that outputs the \textit{composition} of the enclosed \(\ell\) bits and shifts by \(\delta\) positions with each time step. In this context, the composition of a \(q\)-ary vector $\bfx$ specifies the number of occurrences in \(\bfx\) of each symbol in \(\lbrace 0,1,\ldots, q-1\rbrace\). The resulting compositions vector, termed the \emph{read vector}, may also be corrupted by \(t\) substitution errors. By employing graph-theoretic techniques, we deduce that for \(\delta=1\), at least \(\log \log n\) symbols of redundancy are required to correct a single (\(t=1\)) substitution. Finally, for \(\ell \geq 3\),  we exploit some inherent characteristics of read vectors to arrive at an error-correcting code that is of optimal redundancy up to a (small) additive constant 
for this setting. This construction is also found to be optimal for the case of reconstruction from two noisy read vectors.
\end{abstract}

\section{Introduction} \label{sec::intro}

The advent of DNA storage as an encouraging solution to our ever-increasing storage requirements has spurred significant research to develop superior synthesis and sequencing technologies. Among the latter, nanopore sequencing \cite{deamerThree2016, laszloDecoding2014, kasianowiczCharacterization1996} appears to be a strong contender due to low cost, better portability, and support for longer reads. In particular, this sequencing process comprises transmigrating a DNA fragment through a microscopic pore that holds $\ell$ nucleotides at each time instant and measuring the variations in the ionic current, which are influenced by the different nucleotides passing through. However, due to the physical aspects of this process, multiple kinds of distortions corrupt the readout. Firstly, the simultaneous presence of $\ell>1$ nucleotides in the pore makes the observed current dependent on multiple nucleotides instead of just one, thus causing inter-symbol interference (ISI). Next, the passage of the DNA fragment through the pore is often irregular and may involve backtracking or skipping a few nucleotides, thereby leading to duplications or deletions. Furthermore, the measured current is accompanied by random noise, which might result in substitution errors.

Several attempts have been made to develop a faithful mathematical model for the nanopore sequencer. In particular, \cite{maoModels2018} proposed a channel model that embodies the effects of ISI, deletions, and random noise while establishing upper bounds on the capacity of this channel. The authors of \cite{hulettCoding2021} focused on a more deterministic model incorporating ISI and developed an algorithm to compute its capacity. Efficient coding schemes for this abstracted channel were also suggested. More recently, a finite-state Markov channel (FSMC)-based approach was adopted to formulate a model that accounts for ISI, duplications, and noisy measurements \cite{mcbainFiniteState2022a}. 

\begin{figure}[t]
	\scalebox{0.75}{
	\begin{tikzpicture}
	
	\node[align=center] (start) {Nucleotides \\ $\bfx$};
	
	\node[draw, minimum width=1.4cm,
	minimum height=2.6cm, right=0.6cm of start, rounded corners, align=center] (ISI) {ISI \\ $(\ell,\delta)$ };
	
	\node[draw, minimum width=1.6cm,
	minimum height=2.6cm, right=1.2cm of ISI, rounded corners] (sub) {Substitution};
	
	\node[draw, minimum width=1.4cm,
	minimum height=2.6cm, right=0.8cm of sub, rounded corners] (dmc) {DMC};
	
	\node[align=center, right=0.6cm of dmc] (end) {};

	\draw[->] (start) -- (ISI);
	\draw[->] (ISI) --  node[align=center]{$\ell$-mers\\
		$\bfz$} (sub);
	\draw[->] (sub) -- node[below=0.3pt]{$\widehat{\bfz}$} (dmc);
	\draw[->] (dmc) -- (end) node[align=center, right=0.1pt] (h) {Discrete \\ voltage \\ levels \\ $\bfy$};
	
	\end{tikzpicture}}
	\caption{Simplified model of a nanopore sequencer}
    
	\label{fig::ch_model}
\end{figure}
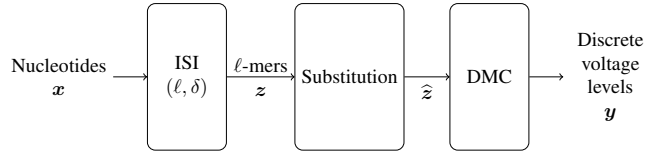

In this work, we adopt a specific variation of the model proposed in \cite{maoModels2018}, which is also interesting owing to its resemblance with the transverse-read channel \cite{cheeCoding2021}, which is relevant to racetrack memories. Expressly, we represent the process of nanopore sequencing as the concatenation of three channels, as depicted in Fig.~\ref{fig::ch_model}. The ISI channel, parameterized by $(\ell, \delta)$, is meant to reflect the dependence of the current variations on the $\ell$ consecutive nucleotides in the pore at any given time. We may view this stage as a sliding window of size $\ell$ passing through an input sequence and shifting by $\delta$ positions after each time instant, thereby producing a sequence of strings of $\ell$ consecutive symbols, or \emph{$\ell$-mers}. Next, the substitution channel captures the effect of random noise by introducing possible substitution errors into the sequence of $\ell$-mers. Finally, this erroneous sequence of $\ell$-mers is converted by a memoryless channel into a sequence of discrete voltage levels according to a deterministic function, specifically the \textit{composition}. 


This work aims to design efficient error-correcting codes for nanopore sequencing. More specifically, as a starting point for future analysis, the aforementioned channel model is treated where at most one substitution occurs and $\delta=1$. 
The problem is stated more formally as follows.

Let $\tr[\ell, \delta]{\bfx}$ represent the channel output for an input $\bfx \in \Sigma_q^n$, given that no substitution affected the $\ell$-mers. 
%
%
Now we seek to find a code $\mathcal{C} \subseteq \Sigma_q^n$ such that for any $\bfx_1, \bfx_2 \in \mathcal{C}$, the Hamming distance between $\tr[\ell, \delta]{\bfx_1}$ and $\tr[\ell, \delta]{\bfx_2}$ strictly exceeds $2$. In other words, one can uniquely deduce the channel input despite ISI and the subsequent occurrence of at most one substitution, provided it belongs to the code $\mathcal{C}$. 


The rest of the manuscript is organized as follows.
We begin by establishing relevant notation and terminology while discussing the underlying properties of read vectors in Section~\ref{sec::prelim}. The results that follow, hold for all $(\ell,1)$-read vectors, where $\ell\geq 3$
. In Section~\ref{sec::lb_red}, we employ graph-theoretic techniques from \cite{chrisnataCorrecting2022} to determine the minimum redundancy required by any code that corrects a single substitution error in an $(\ell,1)$-read vector. Section~\ref{sec::construction} describes a redundancy-optimal 
instantiation of such a code. Subsequently, in Section~\ref{sec::two_reads}, we find that this instantiation is also redundancy-optimal 
when reconstructing $\bfx$ from two distinct noisy copies of $\tr[\ell,\delta]{\bfx}$, each of which has suffered at most $1$ substitution. Concluding remarks concerning future work are offered in Section~\ref{sec::concl}.

\section{Preliminaries} \label{sec::prelim}

\subsection{Notations and Terminology} \label{subsec::not}

In the following, we let $\Sigma_q$ indicate the $q$-ary alphabet $\lbrace0,1,\ldots,q-1\rbrace$. Additionally, $[n]$ is used to denote the set $\lbrace 1,2,\ldots, n\rbrace$. Element-wise modulo operation on a vector, say $\bfy \in \Sigma_q^n$, is represented as
\begin{equation}
\bfy \bmod a\triangleq\big\lparen y_1\bmod a, y_2\bmod a, \ldots, y_n\bmod a \big\rparen. \label{eq::mod_vec}
\end{equation}
For any vector $\bfx=\lparen x_1, \ldots, x_n\rparen$, we refer to its substring $(x_i, x_{i+1}, \ldots, x_j)$ as $\bfx_i^j$. The composition of a vector $\bfx$ is denoted by $c(\bfx) \triangleq 0^{i_0}\ldots (q-1)^{i_{q-1}}$, such that $\bfx$ contains $i_0$~`$0$'s, $i_1$~`$1$'s and so on. We also define the $L_1$-weight of the composition~$c(\bfx)$ as $|c(\bfx)|_1 \triangleq {i_1+2i_2+\cdots+(q-1)i_{q-1}} {=|\bfx|_1}$. This operator may also be applied to a vector of compositions in the same spirit as in~(\ref{eq::mod_vec}). By abuse of notation, when~$n$ is known from the context, we omit from~$c(\bfx)$ any symbol~$x\in \Sigma_q$ such that~$i_x=0$. Further, when convenient, we treat~$c(\bfx)$ as a formal monomial by using expressions of the form $c(\bfx)\cdot (c(\bfy))^{-1}$.

We also extensively use the Hamming distance, which is defined for any two vectors $\bfx, \bfy \in \Sigma^n$, for any alphabet~$\Sigma$, as 
\begin{equation*}
d_H(\bfx, \bfy)=\lvert\lbrace\, i : i\in [n], x_i \neq y_i \,\rbrace\rvert.
\end{equation*}

Throughout this paper, we assume existence of integers $n$, $\ell$, and $\delta$ that satisfy the relation $n+\ell \equiv 0 \pmod{\delta}$. 
\begin{defn} \label{def::rv}
    The $(\ell,\delta)$-\emph{\wread vector} of any ${\bfx \in \Sigma_q^n}$ is of length $(n+\ell)/\delta-1$ and is denoted by
    \begin{equation*}
        \tr[\ell, \delta]{\bfx} \triangleq (c(\bfx_{\delta-\ell+1}^\delta), c(\bfx_{2\delta-\ell+1}^{2\delta}), \ldots, c(\bfx_{n-\delta+1}^{n+\ell-\delta})),
    \end{equation*}
where for any $i\not\in [n]$, we let $x_i=\cnull$, i.e., a null element such that $c(\bfy\; \circ \;\cnull)=c(\bfy)$. $\tri[\ell, \delta]{x}{i}$ is used to denote the $i$-th element of $\tr[\ell, \delta]{\bfx}$, i.e., $\tri[\ell, \delta]{x}{i}=c(\bfx_{i\delta-\ell+1}^{i\delta})$.
\end{defn}


\emph{Remark:} The above definition of an $(\ell, \delta)$-\wread vector appears similar to that of the $(\ell, \delta)$-transverse-read vector introduced in \cite{cheeCoding2021}, except that the $L_1$-weights are replaced by compositions and $\tr[\ell, \delta]{\bfx}$ begins and ends with the {compositions} of substrings $\bfx_{1}^\delta$ and ${\bfx_{n-\delta+1}^n}$ respectively, even though its intermediate elements signify compositions of length-$\ell$ substrings. This is motivated by obtaining a current reading even when the DNA strand has only partially entered the nanopore.


\begin{defn} \label{def::derivative}
    Let $\cR = (c_1,\ldots,c_k)$ where for each $1\leq i\leq k$, $c_k$ is a composition of some vector in~$\Sigma_q^\ell$. Then the \emph{derivative} of~$\cR$ is the length-$(k+1)$ formal-monomial-vector defined as 
    \begin{align*}
        \Delta \triangleq \left\lparen c_1 c_0^{-1}, c_2 c_1^{-1}, \ldots, c_{k+1} c_k^{-1} \right\rparen,
    \end{align*}
    where $c_0,c_{k+1}=\cnull$ are included for uniformity. Observe that the differentiation $\cR\mapsto \Delta$ is invertible.
\end{defn}

\begin{defn}
    For any $\ell, \delta$ where $\ell \geq \delta$, the $i$-{th} \emph{read sub-derivative}, is used to indicate a specific subsequence of the derivative of $\tr[\ell, \delta]{\bfx}$, and is defined for any $\alpha \in \Sigma_{\floor{\frac{\ell}{\delta}}}$ as
    \begin{align*}
        \Delta^\alpha_{\ell,\delta}(\bfx) &\triangleq (\mathcal{R}(\bfx)_{\alpha+1}\cdot\mathcal{R}(\bfx)_{\alpha}^{-1},\mathcal{R}(\bfx)_{\alpha+\floor{\frac{\ell}{\delta}}+1}\cdot \mathcal{R}(\bfx)_{\alpha+\floor{\frac{\ell}{\delta}}}^{-1},  \ldots, \tri{x}{\alpha+k\floor{\frac{\ell}{\delta}}+1}\cdot\tri{x}{\alpha+k\floor{\frac{\ell}{\delta}}}^{-1}) \\
        &= (c(\bfx_{\alpha\delta+1}^{(\alpha+1)\delta})\cdot c(\bfx_{\alpha\delta-\ell+1}^{(\alpha+1)\delta-\ell})^{-1},   \ldots, c(\bfx_{(\alpha+k\floor{\frac{\ell}{\delta}})\delta+1}^{(\alpha+k\floor{\frac{\ell}{\delta}}+1)\delta})\cdot c(\bfx_{(\alpha +k\floor{\frac{\ell}{\delta}})\delta-\ell+1}^{(\alpha+k\floor{\frac{\ell}{\delta}}+1)\delta-\ell})^{-1}), 
    \end{align*}
    where $k=\floor{\frac{n+\ell-(\alpha+1)\delta}{\delta \floor{{\ell}/{\delta}}}}$ and for any $p \not\in [\frac{n+\ell-\delta}{\delta}]$ and ${m\not\in [n]}$, we let $\tri{x}{p}=\cnull$ and $x_m=\cnull$. We let $\Delta^\alpha_{\ell,\delta}(\bfx)_i$ indicate the $i$-th element of $\dtr[\ell,\delta]{\bfx}{\alpha}$, i.e.,
    \begin{equation*}
        \Delta^\alpha_{\ell,\delta}(\bfx)_i= \tri{x}{\alpha+(i-1)\floor{\frac{\ell}{\delta}}+1}\cdot \tri{x}{\alpha+(i-1)\floor{\frac{\ell}{\delta}}}^{-1}.
    \end{equation*}
\end{defn}

When clear from the context, $\ell$ and $\delta$ will be removed from the preceding notations. 


\begin{env_example}
	Consider ${\bfx=(1,2,0,1,2,2)}$. The $(3,1)$-\wread vector of $\bfx$ is thus ${\tr[3,1]{\bfx}=(1,12,012,012,012,12^2,2^2,2)}$. Evidently, $\tri[3,1]{x}{3}=012$, $\dtr[3,1]{\bfx}{0}=(1,\cnull,1^{-1})$, $\dtr[3,1]{\bfx}{1}=(2,\cnull,2^{-1})$ and $\dtr[3,1]{\bfx}{2}=(0,0^{-1} 2,2^{-1})$. \label{eg::read}
\end{env_example}



As mentioned earlier, \cite{cheeCoding2021} investigated a similar model designated as the transverse-read channel in connection with racetrack memories. Therein, the information limit of this channel was derived for different parameters, and several codes enabling unique reconstruction were proposed. Certain error-correcting codes were also presented for $\ell=2$ and $\delta=1$.

\subsection{Properties of the Read Vectors} \label{subsec::tr_prop}

A closer look at the definitions in the last section reveals that not every vector of $\ell$-compositions represents the \wread vector of some $\bfx\in \Sigma_q^n$, i.e., is \emph{valid}. In this section, we first observe which vectors are valid and thereby deduce specific properties that often enable us to detect errors and thereby assist in designing error-correcting constructions of improved redundancies.

\begin{defn} \label{def::cal}
    For~$\bfx\in \Sigma_q^n$ and any~$\alpha \in \Sigma_{\floor{\frac{\ell}{\delta}}}$, let 
    \begin{align*}
        C_{\ell,\delta}^{\alpha}(\bfx)\triangleq (c(\bfx_{\alpha \delta + 1}^{(\alpha + 1) \delta}), \ldots, c(\bfx_{(\alpha + k \lfloor\frac{\ell}{\delta}\rfloor) \delta + 1}^{(\alpha + k \lfloor\frac{\ell}{\delta}\rfloor + 1) \delta})),
    \end{align*}
    where $k=\floor{\frac{n+\ell-(\alpha+1)\delta}{\delta \floor{{\ell}/{\delta}}}}$, be a sequence of compositions.
\end{defn}
Observe that for~$\delta=1$ each~$C_{\ell,\delta}^{\alpha}(\bfx)$ is a subsequence of~$\bfx$, composed of the positions at indices~$i\equiv\alpha+1\pmod{\floor{\frac{\ell}{\delta}}}$, 
and in particular there exists a bijection between~$\Sigma^n$ and the set of $\floor{\frac{\ell}{\delta}}$-tuple of length-$(k+1)$ vectors.


\begin{env_example} \label{eg::subsequences}
    Reconsidering $\bfx=(1,2,0,1,2,2)$ from Example~\ref{eg::read}, we observe that $C^0_{3,1}(\bfx)=(1,1)$, $C^1_{3,1}(\bfx)=(2,2)$ and $C^2_{3,1}(\bfx)=(0,2)$, which are evidently subsequences of $\bfx$. Under $\delta>1$, these transform into composition vectors; for instance $C^0_{4,2}(\bfx)=(12,2^2)$ and $C^1_{4,2}(\bfx)=(01)$.    
\end{env_example}

\begin{lemma} \label{lem::cal-bij}

    Take $\ell,\delta$ satisfying~$\ell\equiv 0 \pmod{\delta}$, and let $\lbrace C_\alpha : \alpha\in \Sigma_{\ell/\delta}\rbrace$ be any $(\ell/\delta)$ arbitrary length-$(k+1)$ vectors of compositions, belonging to vectors in~$\Sigma_q^\delta$, where $k=\floor{\frac{n+\ell-(\alpha+1)\delta}{\ell}}$. Then there exists $\bfx\in \Sigma_q^n$ such that the respective derivatives~$\lbrace \Delta_\alpha : \alpha\in \Sigma_{\ell/\delta}\rbrace$ satisfy $\Delta_\alpha = \dtr[\ell,\delta]{\bfx}{\alpha}$, and~$\bfx$ satisfies $C_\alpha = C_{\ell,\delta}^{\alpha}(\bfx)$ for all~$\alpha\in \Sigma_{\ell/\delta}$. Further, when $\delta=1$ this~$\bfx$ is unique.
\end{lemma}

\begin{IEEEproof}
    Recall that owing to $x_i=\cnull$ for all $i\not\in [n]$, the following holds for any $\alpha \in \frac{\ell}{\delta}$.
    \begin{align*}
        \dtr[\ell,\delta]{\bfx}{\alpha}&=(c(\bfx_{\alpha\delta+1}^{(\alpha+1)\delta}), c(\bfx_{\alpha\delta+\ell+1}^{(\alpha+1)\delta+\ell})\cdot c(\bfx_{\alpha\delta+1}^{(\alpha+1)\delta})^{-1}, \ldots, \\
        &  c(\bfx_{\alpha\delta+k\ell+1}^{(\alpha+1)\delta+k\ell})c(\bfx_{\alpha\delta+(k-1)\ell+1}^{(\alpha+1)\delta+(k-1)\ell})^{-1}, c(\bfx_{\alpha\delta+k\ell+1}^{(\alpha+1)\delta+k\ell})^{-1} ),
    \end{align*}
    where $k=\floor{\frac{n-(\alpha+1)\delta}{\ell}}+1$. Evidently, by left-to-right (or right-to-left) reconstruction, we observe that $C_{\ell,\delta}^{\alpha}(\bfx)$ can be uniquely deduced from $\dtr[\ell,\delta]{\bfx}{\alpha}$. The other direction follows from the observation that $\dtr[\ell,\delta]{\bfx}{\alpha}$ is essentially the derivative of $C_{\ell,\delta}^{\alpha}(\bfx)$, in accordance with Definition~\ref{def::derivative}. 
\end{IEEEproof}

\begin{corollary} 
     If $\ell \equiv 0 \pmod{\delta}$, then for any $\bfx \in \Sigma_q^n$ and $\alpha \in \Sigma_{\frac{\ell}{\delta}}$, the cumulative product of the first $m+1$ elements of $\dtr[\ell,\delta]{\bfx}{\alpha}$ is $c(\bfx_{m\ell+\alpha \delta+1}^{m\ell+(\alpha+1)\delta})$\footnote{Analogous result exists for sum of last $m+1$ elements.}. Thus, $\dtr[\ell,\delta]{\bfx}{\alpha}$ determines $C_{\ell,\delta}^{\alpha}(\bfx)$, which in the special case of $\delta=1$, is effectively $(x_{\alpha+1}, x_{\alpha+\ell+1}, \ldots)$. \label{prop::dtr_sum_first_m}
\end{corollary}

Since $\tr[\ell,\delta]{\bfx}$ is in bijection with the set $\left\lbrace \dtr{\bfx}{\alpha}\right\rbrace_{\alpha\in \Sigma_{\floor{\frac{\ell}{\delta}}}}$, it follows that when~$\ell\equiv 0 \pmod{\delta}$ (and, in particular, when~$\delta=1$) the set of valid read vectors is isomorphic to the set of $\floor{\frac{\ell}{\delta}}$-tuple of length-$(k+1)$ composition-vectors.

\begin{corollary} \label{cor::rec_modq} \label{cl::xij}
    For $\delta=1$ and any $\bfx \in \Sigma_q^n$, let $R(\bfx)$ be either $\tr{\bfx}$ or $|\tr{\bfx}|_1 \bmod q$. Then $\bfx \in \Sigma_q^n$, $\bfx_i^j$ can be uniquely determined, either from
	\begin{enumerate}
		\item $\bfx_{i-\ell+1}^{i-1}$ and $( R(\bfx)_{i}, R(\bfx)_{i+1}, \ldots, R(\bfx)_{j})$; or 
		\item $\bfx_{j+1}^{j+\ell-1}$ and $( R(\bfx)_{i+\ell-1}, R(\bfx)_{i+\ell}, \ldots, R(\bfx)_{j+\ell-1})$,
	\end{enumerate} 
    where for all $k\not\in [n]$, $x_k=\cnull$. Since for $p \in \{1,n\}$, $x_p=R(\bfx)_p$, the first or last $n$ elements of $R(\bfx)$ suffice to reconstruct $\bfx$.
\end{corollary}



\begin{IEEEproof}
We restrict our attention to $R(\bfx)={|\tr{\bfx}|_1 \bmod q}$ since the proof for $\tr{\bfx}$ follows similarly. 
By successively applying the fact that for $i\leq p\leq j$, one can recover $x_p$ from the combined knowledge of $\bfx_{p-\ell+1}^{p-1}$ and $|\tri[\ell,1]{x}{p}|_1 = \sum_{h=p-\ell+1}^{p} x_h \bmod q$, we arrive at the statement of the corollary. The same argument also holds for right-to-left reconstruction.
\end{IEEEproof}


\begin{env_example}
	We wish to reconstruct $\bfx$ given $\tr[3,1]{\bfx}=(1,12,012,012,012,12^2,2^2,2)$ from Example~\ref{eg::read}. Firstly, we observe that $\tri{x}{1}=x_1=1$. Next, $\tri{x}{2}=c(\bfx_1^2)=12$, causing $x_2=2$. Such a left-to-right reconstruction of $\tr{\bfx}$ leads us to ${\bfx=(1,2,0,1,2,2)}$, as in Example~\ref{eg::read}. Similarly, when given $|\tr[3,1]{\bfx}|_1 \bmod 3=(1,0,0,0,0,2,1,2)$, one can infer from Definition~\ref{def::rv} that $x_1=|\tr[3,1]{\bfx}_1|_1 \bmod 3=1$, $x_1+x_2 =|\tr[3,1]{\bfx}_2|_1 \bmod 3$ and so on, thereby leading to ${\bfx=(1,2,0,1,2,2)}$ once again. 
    Right-to-left reconstruction will yield the same result.
\end{env_example}


\begin{lemma} \label{prop::sum_l} \label{lem::sum_d0}  
     For any $\ell, \delta$ such that $\ell \equiv 0 \pmod{\delta}$, and all $\bfx \in \Sigma_q^n$, it holds that $\prod_{i=1}^{\frac{n+\ell}{\delta}-1} \tri{\bfx}{i}=   \big(c(\bfx)\big)^{\ell/\delta}$. Further, $\prod_{i=1}^{\lfloor\frac{n-(\alpha+1)\delta}{\ell}\rfloor+2} \dtr[\ell,\delta]{\bfx}{\alpha}_i=c(\cnull)$ for any $\alpha\in\Sigma_{\ell/\delta}$.
\end{lemma}

\begin{IEEEproof}
    Observe that for all $\alpha \in \Sigma_{\ell/\delta}$, we have
    \begin{align}
        \prod_{i=0}^{\big\lfloor\frac{n-(\alpha+2)\delta}{\ell}\big\rfloor+1}\tri{\bfx}{\alpha+i\frac{\ell}{\delta}+1}&=c(\bfx). \label{eq::prod_r}
    \end{align}
    This naturally leads us to 
    \begin{align*}
        \prod_{i=1}^{\frac{n+\ell}{\delta}-1}\tri{\bfx}{i}=\prod_{\alpha=0}^{\ell/\delta-1}\prod_{i=0}^{\lfloor\frac{n-(\alpha+2)\delta}{\ell}\rfloor+1}\tri{\bfx}{\alpha+i\frac{\ell}{\delta}+1}&=c(\bfx)^{\ell/\delta}.
    \end{align*}
    While one can arrive at $\prod_{i=1}^{\lfloor\frac{n-(\alpha+1)\delta}{\ell}\rfloor+2} \dtr{\bfx}{\alpha}_i=c(\cnull)$ directly from the definition, we may also use (\ref{eq::prod_r}) to prove this as follows.
    \begin{align*}
        \prod_{i=1}^{\lfloor\frac{n-(\alpha+1)\delta}{\ell}\rfloor+2} \dtr{\bfx}{\alpha}_i 
        &= \prod_{i=0}^{\lfloor\frac{n-(\alpha+1)\delta}{\ell}\rfloor+1} \tri{\bfx}{\alpha+i\frac{\ell}{\delta}+1} \cdot \tri{x}{\alpha+i\frac{\ell}{\delta}}^{-1} \\
        &= \Big( \prod_{i=0}^{\floor{\frac{n-(\alpha+1)\delta}{\ell}}+1} \tri{\bfx}{\alpha+i\frac{\ell}{\delta}+1} \Big)\Big(\prod_{i=0}^{\floor{\frac{n-(\alpha+1)\delta}{\ell}}+1} \tri{x}{\alpha+i\frac{\ell}{\delta}}\Big)^{-1} \\
        &=c(\bfx)\cdot c(\bfx)^{-1} =\cnull,
    \end{align*}
    where we let $\tri{x}{p}=\phi$ for all $p\not\in [\frac{n+\ell-\delta}{\delta}]$.
\end{IEEEproof}

Another important consequence of the aforementioned properties is stated below. 

\begin{theorem}\label{th::no_dh1}
	When $\ell>1$ and $\delta=1$, for any two distinct $\bfx, \bfy \in \Sigma_q^n$, $d_H(\tr{\bfx}, \tr{\bfy}) \geq 2$. 
\end{theorem}

\begin{IEEEproof}
	Assume that $d_H(\tr{\bfx}, \tr{\bfy})=1$, and let $i$ denote the index where $\tr{\bfx}$ and $\tr{\bfy}$ differ, i.e., ${\tri{x}{i} \neq \tri{y}{i}}$. From Lemma~\ref{prop::sum_l}, we infer that
	\begin{align*}
    \big(c(\bfx)\cdot c(\bfy)^{-1} \big)^{\ell} = \prod_{j=1}^{\mathclap{\frac{n+\ell}{\delta}-1}}
	\tri{\bfx}{j}\cdot \tri{\bfy}{j}^{-1} =	\tri{x}{i}\cdot \tri{y}{i}^{-1}. 
	\end{align*}
	
	Since the preceding equation suggests that each positive and negative degree should be divisible by $\ell$, and we know that the sum of degrees in each of $\tri{x}{i}$ and $\tri{y}{i}$ must be $\ell$, the only possibility involves $\tri{\bfx}{i} \cdot \tri{\bfy}{i}^{-1}=a^\ell b^{-\ell}$ for some $a,b\in \Sigma_q$, $a\neq b$. 
 
    However, denoting $\alpha \triangleq (i-1) \bmod \ell$ it also follows that $\dtr{\bfx}{\alpha}$ and $\dtr{\bfy}{\alpha}$ differ in a unique index, at which $\tri{\bfx}{i} \cdot \tri{\bfx}{i-1}^{-1} = (\tri{\bfy}{i} a^\ell b^{-\ell}) \cdot \tri{\bfy}{i-1}^{-1} \neq \tri{\bfy}{i} \cdot \tri{\bfy}{i-1}^{-1}$. Hence, by Lemma~\ref{lem::sum_d0} 
    \begin{align*}
        c(\phi) &= 
        \prod_{i=1}^{\floor{\frac{n-(\alpha+1)\delta}{\ell}}+2} \dtr{\bfx}{\alpha}_i \\
        &= a^\ell b^{-\ell} \prod_{i=1}^{\floor{\frac{n-(\alpha+1)\delta}{\ell}}+2} \dtr{\bfy}{\alpha}_i 
        = a^\ell b^{-\ell} c(\phi), 
    \end{align*} 
    in contradiction. 
\end{IEEEproof}

\subsection{Error Model} \label{subsec::err_mod}

Similar to \cite{cheeCoding2021}, we study the occurrence of substitution errors in \wread vectors and design suitable error-correcting constructions.  To suitably define what constitutes an error-correcting construction in our framework, we first define the set of vectors that may result from at most $t$ substitutions on a vector $\bfu \in \Sigma^n$, for any alphabet~$\Sigma$, as
\begin{equation}
    B_{t}(\bfu) \triangleq \{ \bfv \in \Sigma^n : d_H\big(\bfu, \bfv \big) \leq t\}. \label{eq::err_ball}
\end{equation} 
In our application we will only be interested in~$B_t(\tr{\bfx})$, for some~$\bfx\in \Sigma_q^n$. 



\begin{defn}
    A code $\mathcal{C}$ is said to be a $t$-substitution $(\ell,\delta;M)$-read code if for any two distinct $\bfx,\bfy \in \Sigma_q^n$, it holds that $|B_{t}(\tr[\ell,\delta]{\bfx}) \cap B_{t}(\tr[\ell, \delta]{\bfy})| < M$.
\end{defn}

In words, $\mathcal{C}$ is a $t$-substitution $(\ell,\delta;M)$-read code if obtaining any $M$ distinct noisy versions of any codeword, where at most~$t$ substitutions occur in each version, allows one to uniquely reconstruct that codeword.

This work focuses on the case when $\delta=1$ and $t=1$. To this end, we seek to find a code that can correct a single substitution error in the \wread vectors of its constituent codewords, i.e., a single-substitution $(\ell,1)$-\wread code. 
In the upcoming sections, we endeavor to derive an upper bound on the cardinality of any such 
code, and subsequently propose an optimal instantiation of the same.

\section{Minimum Redundancy of 
Single-substitution \texorpdfstring{$(\ell,1;1)$-\wread}{(l,1)-read} codes} \label{sec::lb_red}

To establish a lower bound on the redundancy required by a single-substitution $(\ell,1;1)$-\wread code, we first attempt to characterize the relationship between any two non-binary vectors $\bfx, \bfy \in \Sigma_q^n$
, that might be confusable after a single substitution in their respective \wread vectors.

\subsection{Characterization of Confusable Read Vectors} \label{sec::char_tr}

 To proceed in this direction, we first note from Theorem~\ref{th::no_dh1} that there exists no two distinct vectors ${\bfx, \bfy \in \Sigma_q^n}$ that satisfy ${d_H(\tr{\bfx}, \tr{\bfy})=1}$ for any ${\ell>1}$. Thus, we attempt to ascertain the conditions under which $d_H(\tr{\bfx}, \tr{\bfy})=2$ may occur.

 \begin{lemma}
 	For $\ell \geq 3$, any two distinct vectors ${\bfx, \bfy \in \Sigma_q^n}$ satisfy $d_H(\tr{\bfx}, \tr{\bfy})=2$ if and only if there exist distinct $i,j \in [n+\ell-1]$, for which $\tri{x}{i}\cdot \tri{y}{i}^{-1} = \tri{x}{j}^{-1} \cdot \tri{y}{j} \linebreak[0]\neq c(\cnull)$ and $j\equiv i \pmod{\ell}$. 
     \label{lem::dh2_1n1}
 \end{lemma}

 \begin{IEEEproof}
 	Let $i<j$ represent the indices at which $\tr{\bfx}$ and $\tr{\bfy}$ differ, i.e., $\tri{\bfx}{i} \neq \tri{\bfy}{i}$ and $\tri{\bfx}{j} \neq \tri{\bfy}{j}$. As $(\tr{\bfx}_1^{i-1}, \tr{\bfx}_{j+1}^{n-\ell+1})=(\tr{\bfy}_1^{i-1}, \tr{\bfy}_{j+1}^{n-\ell+1})$, we may infer from Corollary~\ref{cl::xij} that $\bfx_1^{i-1}=\bfy_1^{i-1}$ and $\bfx_{j-\ell+2}^n=\bfy_{j-\ell+2}^n$. As a consequence, we obtain $x_{i}\cdot y_{i}^{-1} = \tri{x}{i}\cdot \tri{y}{i}^{-1} \neq c(\cnull)$, i.e., $x_i \neq y_i$. 
	
 	Similarly,  $x_{j-\ell+1}\cdot y_{j-\ell+1}^{-1} = \tri{x}{j}\cdot \tri{y}{j}^{-1} \neq c(\cnull)$. On account of Lemma~\ref{prop::sum_l}, we also have
    \begin{align*}
        (x_{i}\cdot y_{i}^{-1}) (x_{j-\ell+1}\cdot y_{j-\ell+1}^{-1}) &= \prod_{i=1}^{n+\ell-1} \tri{\bfx}{i} \cdot \tri{\bfy}{i}^{-1} \\
        &= c(\bfx)^{\ell} c(\bfy)^{-\ell}, 
    \end{align*}
     hence the degree in $(x_{i}\cdot y_{i}^{-1}) (x_{j-\ell+1}\cdot y_{j-\ell+1}^{-1})$ of each symbol in~$\Sigma_q$ is a multiple of~$\ell$. Since $\ell \geq 3$, it follows that $(x_{i}\cdot y_{i}^{-1}) (x_{j-\ell+1}\cdot y_{j-\ell+1}^{-1}) = c(\cnull)$. Because $x_i\neq y_i$, we have $x_i = y_{j-\ell+1}$ and $y_i = x_{j-\ell+1}$ (i.e., $c(\bfx)=c(\bfy)$), or, put differently, 
     $\tri{x}{j}\cdot \tri{y}{j}^{-1} =x_i^{-1}y_i = \tri{x}{i}^{-1}\cdot \tri{y}{i}$.

    Finally, if $j\not\equiv i,i+1\pmod{\ell}$ then under Lemma~\ref{lem::sum_d0} we observe (denoting $\alpha \triangleq i \bmod \ell$) 
    \begin{align*}
        c(\cnull) &= \prod_k \dtr{\bfx}{\alpha}_k \\
        &= \Big\lparen \prod_k \dtr{\bfy}{\alpha}_k \Big\rparen 
        (\tri{x}{i}^{-1} \tri{y}{i}) \\
        &= \tri{x}{i}^{-1} \tri{y}{i}, 
    \end{align*}
    in contradiction. Similarly, if $j\equiv i+1\pmod{\ell}$ then 
    \begin{align*}
        c(\cnull) &= \prod_k \dtr{\bfx}{\alpha}_k \\
        &= \Big\lparen \prod_k \dtr{\bfy}{\alpha}_k \Big\rparen 
        (\tri{x}{i}^{-1} \tri{y}{i}) (\tri{x}{j} \tri{y}{j}^{-1}) \\
        &= \tri{x}{i}^{-2} \tri{y}{i}^2, 
    \end{align*}
    again in contradiction. Hence, $i\equiv j\pmod{\ell}$, concluding the proof.
    
%
\end{IEEEproof}

Further inspection reveals how two binary vectors with confusable \wread vectors are related.

\begin{lemma}
	For $\ell \geq 3$%
    , any two vectors $\bfx,\bfy \in \Sigma_q^n$ that satisfy $d_H(\tr{\bfx}, \tr{\bfy})=2$, i.e., $\tri{x}{i}\cdot \tri{y}{i}^{-1} = \tri{x}{j}^{-1} \cdot \tri{y}{j}$ for some $i,j \in [n+\ell-1]$ such that $i<j$, it must hold that $(y_i, y_{i+1})=(x_{i+1}, x_{i})$.\label{lem::xy1001_p1}
\end{lemma}

\begin{IEEEproof}
	From Corollary~\ref{cl::xij} and $\tr{\bfx}_1^{i-1}=\tr{\bfy}_1^{i-1}$, we infer that $\bfx_1^{i-1}= \bfy_1^{i-1}$ and $\bfx_{j-\ell+2}^n=\bfy_{j-\ell+2}^n$. It directly follows from $\tri{x}{i}\cdot \tri{y}{i}^{-1}=x_{i}\cdot y_{i}^{-1} \neq c(\cnull)$.
	
	Note that Lemma~\ref{lem::dh2_1n1} suggests $j-i\geq \ell \geq 3$%
    . Thus, we must have $\tri{x}{i+1}=\tri{y}{i+1}$, or equivalently, $\tri{x}{i+1}\cdot \tri{y}{i+1}^{-1}=c(\cnull)$. Since $\bfx_{i-\ell+2}^{i-1}=\bfy_{i-\ell+2}^{i-1}$, the preceding requirement essentially translates to $c(\bfx_i^{i+1}) \cdot c(\bfy_i^{i+1})^{-1} =c(\cnull)$. Since $x_i \neq y_i$, we conclude that $x_{i+1}=y_i$ and $y_{i+1}=x_i$.
\end{IEEEproof}

\begin{env_example}
    For $\bfx=(1,2,0,1,2,2)$ and ${\bfy=(2,1,0,2,1,2)}$, we have ${\tr[3,1]{\bfx}=(1,12,012,012,012,12^2,2^2,2)}$ and $\tr[3,1]{\bfy} = (2,12,012,012,012,12^2,12,2)$ respectively. Evidently, $d_H(\tr{\bfx}, \tr{\bfy})=2$. If $i=1$ and $j=7$, then $\tri{x}{i}\cdot \tri{y}{i}^{-1}=\tri{y}{j}\cdot \tri{x}{j}^{-1}$ and $j\equiv i \pmod{\ell}$, in keeping with Lemma~\ref{lem::dh2_1n1}. Also, as suggested by Lemma~\ref{lem::xy1001_p1}, it holds that $(x_i,x_{i+1})=(1,2)=(y_{i+1},y_i)$. \label{eg::xy1001_p1}
\end{env_example}

\begin{lemma}
    For $\ell \geq 3$, consider two vectors $\bfx,\bfy \in \Sigma_q^n$, such that for some ${i,j \in [n+\ell-1]}$, $i<j$, $\tri{x}{i}\cdot \tri{y}{i}^{-1}=\tri{x}{j}^{-1}\cdot \tri{y}{j}$, and for all ${p \not\in \{i,j\}}$, $\tri{x}{p}=\tri{y}{p}$. Assume for some $t\geq i$ that $\bfx_{t-\ell+2}^{t-1}=\bfy_{t-\ell+2}^{t-1}$, $(x_t, x_{t+1})=(y_{t+1}, y_t)$. Then, one of the following conditions will hold.
    \begin{enumerate}
        \item $\bfx_{t+2}^n=\bfy_{t+2}^n$ and $j=t+\ell$; or
        \item $\bfx_{t+2}^{t+\ell-1}=\bfy_{t+2}^{t+\ell-1}$, $(x_{t+\ell}, x_{t+\ell+1})=(x_{t}, x_{t+1})$, $(y_{t+\ell}, y_{t+\ell+1})=(y_{t}, y_{t+1})$ and $j> t+\ell+1$.
    \end{enumerate} \label{lem::xy1001_p2}
\end{lemma}
\begin{IEEEproof}
    Let $m$ indicate the smallest index strictly greater than $t+1$ for which $x_m \neq y_m$; if $\bfx_{t+2}^n = \bfy_{t+2}^n$, consider ${m = \infty}$. 
    Further recall throughout the proof that from $\tr{\bfx}_{p}=\tr{\bfy}_{p}$ for $j< p <n+\ell$ and Corollary~\ref{cl::xij}, it follows that $\bfx_{j-\ell+2}^n=\bfy_{j-\ell+2}^n$.

    We start by noting that $m\geq t+\ell$; indeed, if $m < t+\ell$ then $\tri{y}{m}\cdot \tri{x}{m}^{-1}=y_m\cdot x_m^{-1} \neq c(\cnull)$, hence $j=m$ but 
    $\bfx_{j-\ell+2}^n=\bfy_{j-\ell+2}^n$ contradicts $x_{t+1}\neq y_{t+1}$. We continue this proof by cases.

    \emph{Case 1)} If $m>t+\ell$, we deduce that $\tri{y}{t+\ell}\cdot \tri{x}{t+\ell}^{-1} =y_{t+1}\cdot x_{t+1}^{-1} \neq c(\cnull)$. Thus, $j=t+\ell$, implying $\bfx_{t+2}^n=\bfy_{t+2}^n$ (in particular, this case is only possible when $m = \infty$).
    
    \emph{Case 2)} If $m=t+\ell$, then $\bfx_{j-\ell+2}^n=\bfy_{j-\ell+2}^n$ implies that $j\geq t+2\ell-1 > t+\ell+1$.
    
    Observe that $\tri{y}{t+\ell}\cdot \tri{x}{t+\ell}^{-1} = (y_{t+\ell}y_{t+1}) \cdot (x_{t+\ell} x_{t+1})^{-1}=c(\cnull)$, and hence $(x_{t+\ell}, y_{t+\ell}) = (y_{t+1}, x_{t+1})$. In turn, $\tri{y}{t+\ell+1}\cdot \tri{x}{t+\ell+1}^{-1} = c(\bfy_{t+\ell}^{t+\ell+1})\cdot c(\bfx_{t+\ell}^{t+\ell+1})^{-1}=c(\cnull)$ now implies $(x_{t+\ell+1}, y_{t+\ell+1}) = (y_{t+\ell}, x_{t+\ell})$.
\end{IEEEproof}

\begin{env_example}
	To demonstrate the implications of Lemma~\ref{lem::xy1001_p2}, we refer back to Example~\ref{eg::xy1001_p1} and note that when $t\in \{1,4\}$, we have $\bfx_{t-\ell+2}^{t-1}=\bfy_{t-\ell+2}^{t-1}$, $\bfx_t^{t+1}=(1,2)$ and $\bfy_t^{t+1}=(2,1)$. As before, let $j$ denote the last index where $\tri{y}{j}\cdot \tri{x}{j}^{-1} \neq c(\cnull)$. Now when $t=4$, it holds that $\bfx_{t+2}^n=\bfy_{t+2}^n$ and ${j=t+\ell}$. On the other hand, when $t=1$, we observe that $\bfx_{t+2}^{t+\ell-1}=\bfy_{t+2}^{t+\ell-1}$, $\bfx_{t+\ell}^{t+\ell+1}=(1,2)$ and $\bfy_{t+\ell}^{t+\ell+1}=(2,1)$ and $j=t+2\ell$.
\end{env_example}

Upon successive applications of Lemma~\ref{lem::xy1001_p2} in conjunction with Lemma~\ref{lem::dh2_1n1} and Lemma~\ref{lem::xy1001_p1}, we arrive at the following theorem.

\begin{theorem}\label{th::xy1001}
    For $\ell \geq 3$ and any $\bfx, \bfy \in \Sigma_q^n$, the following statements are equivalent:
    \begin{enumerate}
        \item $d_H(\tr{\bfx}, \tr{\bfy})=2$.
        \item There exist distinct $i,j \in [n+\ell-1]$, $j\equiv i \pmod{\ell}$, such that $\tri{x}{i}\cdot \tri{y}{i}^{-1}=\tri{x}{j}^{-1} \cdot \tri{y}{j} \neq c(\cnull)$ and $\tri{x}{r}=\tri{y}{r}$ for all $r\not\in \bracenv*{i,j}$.
        \item There exist $p\geq 1$ and $i\in [n - (p-1) \ell - 1]$ such that for all $m \in \Sigma_{p}$ it holds that ${\bfx_{i+m\ell}^{i+m\ell+1} = (a,b)}$, ${\bfy_{i+m\ell}^{i+m\ell+1} = (b,a)}$ (or vice versa) where $a,b \in \Sigma_q$ and $a\neq b$, and $x_r = y_r$ for all $r\not\in \bigcup_{m\in \Sigma_{p}} \lbrace i+m\ell, i+m\ell+1\rbrace$.
    \end{enumerate}
    Further, if these conditions hold, then $j = i + p\ell$ in the above notation.
\end{theorem}

\subsection{An Upper Bound on the Code Size} \label{subsec::cq}

We derive a lower bound on the redundancy required by a single-substitution $(\ell,1;1)$-\wread code by adopting the approach employed in \cite{chrisnataCorrecting2022}. More precisely, we consider a graph $\mathcal{G}(n)$ containing vertices corresponding to all vectors in $\Sigma_q^n$. Any two vertices in $\mathcal{G}(n)$ that signify two distinct binary vectors, say $\bfx, \bfy \in \Sigma_q^n$, are considered to be adjacent if and only if $d_H(\tr{\bfx}, \tr{\bfy})=2$. Therefore, any \emph{independent set} (i.e., a subset of vertices of $\mathcal{G}(n)$, wherein no two vertices are adjacent) is a $1$-substitution $(\ell,1;1)$-read code.


\begin{defn}
	A \textbf{clique cover} $\mathcal{Q}$ is a collection of cliques in a graph $\mathcal{G}$, such that every vertex in $\mathcal{G}$ belongs to at least one clique in $\mathcal{Q}$.
\end{defn}

The following graph-theoretic result is well-known~\cite{knuthSandwich1994}.
\begin{theorem}
	If $\mathcal{Q}$ is a clique cover, then the size of any independent set is at most $|\mathcal{Q}|$. \label{th::cq_cover}
\end{theorem}

For the remainder of this section, we seek to define a clique cover $\mathcal{Q}$ by utilizing Theorem~\ref{th::xy1001}. By Theorem~\ref{th::cq_cover}, the size of such a clique cover will serve as an upper bound on the cardinality of a $1$-substitution $(\ell,1;1)$-\wread code.

\begin{defn}
    Let $\mathcal{G}'(n)$ be the graph whose vertices are all vectors in $\Sigma_q^n$, and an edge connects $\bfx,\bfy\in \Sigma_q^n$ if and only if $\{\bfx,\bfy\} = \{\bfu \circ (a b)^j\circ \bfv, \bfu \circ (b a)^j\circ \bfv\}$, for some $j$, sub-strings $\bfu,\bfv$ and $a,b \in \Sigma_q$ where $a\neq b$. \label{def::gp}
\end{defn}

Observe that when $q=2$, the preceding definition is identical to that in \cite[Sec.~IV]{chrisnataCorrecting2022}. 

Our method of proof would be to pull back a clique cover from $\mathcal{G}'$ based on the non-binary extension of \cite[Lem.~7]{chrisnataCorrecting2022}, i.e., Lemma~\ref{lem::ccq}, into $\mathcal{G}$. To do that, we have the following definition:

\begin{defn}\label{defn:pi_p}
    For a positive integer~$p$, define a permutation~$\pi_p$ on $\Sigma_q^n$ as follows. For all $\bfx\in \Sigma_q^n$, arrange the coordinates of $\bfx_1^{p\ell \floor{n/(p\ell)}}$ in a matrix $X\in \Sigma^{p\floor{n/(p\ell)}\times \ell}$, by row (first fill the first row from left to right, then the next, etc.). 
    Next, partition $X$ into sub-matrices of dimension $p\times 2$ (if $\ell$ is odd, we ignore $X$'s right-most column). 
    Finally, going through each sub-matrix (from left to right, and then top to bottom), we concatenate its \emph{rows} to obtain $\pi_p(\bfx)$ (where unused coordinates from~$\bfx$ are appended arbitrarily).
    
    More precisely, for all $0\leq i < \floor{\frac{n}{p \ell}}$, $0\leq j < \floor{\frac{\ell}{2}}$ and $0\leq k < p$ denote 
    \[
    \bfx^{(i,j,k)} \triangleq \linebreak x_{(i p+k)\ell+2j+1} x_{(i p+k)\ell+2j+2};
    \]
    then 
    \[
    \bfx^{(i,j)} \triangleq \bfx^{(i,j,0)}\circ \cdots\circ \bfx^{(i,j,p-1)}
    \]
    and 
    \[
    \bfx^{(i)} \triangleq \bfx^{(i,0)}\circ \cdots\circ \bfx^{(i,\floor{\ell/2}-1)}.
    \]
    Then $\pi_p(\bfx) = \bfx^{(0)}\circ \cdots\circ \bfx^{(\floor{n/p \ell}-1)} \circ \widetilde{\bfx}$, where $\widetilde{\bfx}$ is composed of all coordinates of $\bfx$ not previously 
    included.
\end{defn}

\begin{env_example} \label{eg::xy_pi}
	For $\bfx=(1,2,0,1,2,2)$ and ${\bfy=(2,1,0,2,1,2)}$ it holds that $d_H(\tr[3,1]{\bfx}, \tr[3,1]{\bfy})=2$. To obtain $\pi_p(\bfx)$ and $\pi_p(\bfy)$ for $p=2$, note that
	\begin{equation*}
		X=\begin{bmatrix}
		1 & 2 & 0 \\
		1 & 2 & 2
		\end{bmatrix},\; Y=\begin{bmatrix}
		2 & 1 & 0 \\
		2 & 1 & 2
		\end{bmatrix}.
	\end{equation*}
	
	Since $\ell$ is odd, we ignore the last column in $X$ and $Y$ and partition the respective results into $2 \times 2$ sub-matrices to ultimately obtain $\pi_p(\bfx)=(1,2,1,2,0,2)$ and $\pi_p(\bfy)=(2,1,2,1,0,2)$ (here, unused coordinates were appended in the order of their indices).
\end{env_example}

\begin{defn}
	For a positive integer~$p$, let 
	\[
	\Lambda_{p,a,b} \triangleq \mathset*{(\bfv)^j (\bft)^{p-j}}{j\in [p], \bracenv*{\bfv,\bft} = \bracenv*{ab,ba}},
	\]
	where $\bfv^0=\bft^0$ is the empty word, and $\widetilde{\Lambda}_{p,a,b} \triangleq \Sigma_q^{2p}\setminus \Lambda_{p,a,b}$.
	Further, let 
	\begin{align*}
	\Gamma \triangleq \Big\lbrace (\bfu,\bfw,a,b) : \;&i\in [m], \bfu\in \widetilde{\Lambda}_{p,a,b}^{i-1}, \bfw\in \Sigma_q^{2p(m-i)}, \\
    & a,b \in \Sigma_q, a\neq b \Big\rbrace,
	\end{align*}
	where $m = \floor{\frac{\ell}{2}} \floor{\frac{n}{p\ell}}$, and $\widetilde{\Lambda}_p^0$ is the singleton containing the empty word. Then, for all $\boldsymbol{\gamma}=(\bfu,\bfw,a,b)\in \Gamma$ define 
	\begin{align*}
		Q_{\gamma}^{(0)} &\triangleq \mathset*{\bfu (ab)^h (ba)^{p-h} \bfw}{h\in [p]}, \\
		Q_{\gamma}^{(1)} &\triangleq \mathset*{\bfu (ba)^h (ab)^{p-h} \bfw}{h\in [p]}.
	\end{align*}
	Finally, let 
	\[
	\mathcal{Q}(m,p) \triangleq \mathset*{\bracenv*{\bfx}}{\bfx\in \widetilde{\Lambda}_p^m} 
	\cup \mathset*{Q_\gamma^{(0)},Q_\gamma^{(1)}}{\gamma\in \Gamma},
	\]
    where $\widetilde{\Lambda}_p=\Sigma_q^{2p}\setminus \cup_{\substack{a,b \in \Sigma_q \\ a\neq b}}\Lambda_{p,a,b}$.
\end{defn}


\begin{env_example}
    For $p=2$, $a=1$ and $b=2$, we obtain $\Lambda_{p,a,b}=\{ (1,2,2,1), (1,2,1,2), (2,1,1,2), (2,1,2,1)\}$.
    Revisiting Example~\ref{eg::xy_pi}, we observe that for $\boldsymbol{\gamma}=\big( \bfu, \bfw, 1, 2 \big) \in \Gamma$, where $\bfu=\widetilde{\Lambda}_{p,a,b}^0$ and $\bfw=(2)$,
    \begin{align*}
        Q_{\gamma}^{(0)} &=\{  (1,2,0,2,1,2), \;\bfx= (1,2,0,1,2,2)\}, \\
        Q_{\gamma}^{(1)} &=\{  (2,1,0,1,2,2), \;\bfy=(2,1,0,2,1,2)\}.
    \end{align*}
    It follows from Theorem~\ref{th::xy1001} that $Q_{\gamma}^{(0)} \cup Q_{\gamma}^{(1)}$ forms a clique.
\end{env_example}

\begin{lemma}
	$\mathcal{Q}(m,p)$ is a clique-cover of $\mathcal{G}'(2p m)$, where ${m = \floor{\frac{\ell}{2}} \floor{\frac{n}{p\ell}}}$. \label{lem::ccq}
\end{lemma}

This is the non-binary analogue of \cite[Lemma 7]{chrisnataCorrecting2022} and the proof is relegated to the appendix.

\begin{theorem}
	Let 
	\[
	\mathcal{Q}_p \triangleq \mathset*{\pi_p^{-1}(Q\times \bracenv*{\bfz})}{Q\in \mathcal{Q}(m,p), \bfz\in \Sigma_q^{n-2p m}}, 
	\]
	where $\pi_p^{-1}(A) \triangleq \mathset*{\bfu\in \Sigma_q^n}{\pi_p(\bfu)\in A}$. Then, $\mathcal{Q}_p$ is a clique-cover in $\mathcal{G}(n)$. \label{th::cqc_specified}
\end{theorem}
\begin{IEEEproof}
First, observe that it readily follows from $\bigcup \mathcal{Q}(m,p) = \Sigma_q^{2p m}$ that $\bigcup \mathcal{Q}_p = \Sigma_q^n$. It is therefore left to prove that every element of $\mathcal{Q}_p$ is a clique of $\mathcal{G}(n)$.

Next, observe for all $Q\in \mathcal{Q}(m,p)$ and $\bfz\in \Sigma_q^{n-2p m}$ that either $Q$ is a singleton, or all elements $\bfy\in Q\times\bracenv*{\bfz}$ agree on all coordinates~$y_k$ except $2(i-1)p < k \leq 2ip$ for some $i\in [m]$, and $\bfy_{2(i-1)p}^{2ip}\in \bracenv*{(ab)^h (ba)^{p-h},(ba)^h (ab)^{p-h}}$ for some $h\in [p]$ and $a,b \in \Sigma_q$ where $a\neq b$. 
That is, either $\pi_p^{-1}(Q\times \bracenv*{\bfz})$ is a singleton, or all elements $\bfx\in \pi_p^{-1}(Q\times \bracenv*{\bfz})$ agree on all coordinates except, in the notation of Definition~\ref{defn:pi_p}, $\bfx^{(i,j)}$ for some $0\leq i < \floor{\frac{n}{p \ell}}$, $0\leq j < \floor{\frac{\ell}{2}}$, and $\bfx^{(i,j)} \in \bracenv*{(ab)^h (ba)^{p-h},(ba)^h (ab)^{p-h}}$ for some $h\in [p]$. That is, $\bfx^{(i,j,k)} = ab$ ($ba$) for all $0\leq k<h$, and $\bfx^{(i,j,k)} = ba$ (respectively, $ab$) for all $h\leq k<p$. By Theorem~\ref{th::xy1001}, it holds that $d_H(\tr{\bfx_1}, \tr{\bfx_2}) = 2$ for all $\bfx_1,\bfx_2\in \pi_p^{-1}(Q\times \bracenv*{\bfz})$.
\end{IEEEproof}

Finally, we can obtain a lower bound on the redundancy of a single-substitution $(\ell,1;1)$-\wread code from the following result.

\begin{lemma}
    \begin{IEEEeqnarray*}{+rCl+x*}
        \abs*{\cQ_p}
        &=& q^{n} \bigg[\parenv*{1 - \binom{q}{2} \frac{2p}{q^{2p}}}^m \>+  \frac{1}{p} \binom{q}{2} 
        \parenv[\bigg]{1 - \parenv[\Big]{1 - \frac{2p}{q^{2p}}}^m}\bigg],
    \end{IEEEeqnarray*}
	where $m = \floor{\frac{\ell}{2}} \floor{\frac{n}{p\ell}}$.
\end{lemma}
\begin{IEEEproof}
    Since the number of singletons is given by
    \begin{align*}
        |\widetilde{\Lambda}_p^m| =\bigg(q^{2p} - \binom{q}{2}2p \bigg)^m,
    \end{align*}
    while the number of cliques of size $p$ evaluates to
    \begin{align*}
        2|\Gamma|&=2\binom{q}{2} \sum_{i=1}^{m} |\widetilde{\Lambda}_{p,a,b}|^{i-1}\cdot q^{2p(m-i)} \\
        &=2\binom{q}{2} \sum_{i=1}^{m} (q^{2p} -2p )^{i-1}q^{2p(m-i)} \\
        &= 2 q^{2p(m-1)} \binom{q}{2} \sum_{i=1}^{m} 
        \parenv*{1 - \frac{2p}{q^{2p}}}^{i-1} \\
        &= q^{2pm} \frac{1}{p} \binom{q}{2} 
        \parenv[\bigg]{1 - \parenv[\Big]{1 - \frac{2p}{q^{2p}}}^m}. 
    \end{align*}
    Hence, 
    \begin{IEEEeqnarray*}{+rCl+x*}
        \abs*{\mathcal{Q}(m,p)} 
        &=& q^{2p m} \bigg[\parenv*{1 - \binom{q}{2} \frac{2p}{q^{2p}}}^m \>+  \frac{1}{p} \binom{q}{2} 
        \parenv[\bigg]{1 - \parenv[\Big]{1 - \frac{2p}{q^{2p}}}^m}\bigg],
    \end{IEEEeqnarray*}
    and the claim follows.
\end{IEEEproof}

By using $\Big(1-\frac{2p}{q^{2p}} \Big) \geq \Big(1-\binom{q}{2}\frac{2p}{q^{2p}} \Big)$, it readily follows that for any positive integer~$p$, 
\begin{IEEEeqnarray*}{+rCl+x*}
    \log_q\abs*{\mathcal{Q}_p} &\leq& n - \log_q(p) \>+  \log_q\parenv*{p \parenv[\Big]{1 - \frac{2p}{q^{2p}}}^m + \binom{q}{2}}.
\end{IEEEeqnarray*}
Based on $m\geq \floor{\frac{n}{2p}}-\floor{\frac{\ell}{2}}$ we may further bound 
\begin{IEEEeqnarray*}{+rCl+x*}
    \log_q\abs*{\mathcal{Q}_p} &\leq& n - \log_q(p) \>+  \log_q\parenv*{p \parenv[\Big]{1 - \frac{2p}{q^{2p}}}^{\floor{n/2p}-\floor{\ell/2}} + \binom{q}{2}}.
\end{IEEEeqnarray*}
By employing the non-binary extension of \cite[Lemma 9]{chrisnataCorrecting2022}, as stated in Appendix~\ref{app::o1}, we find that letting ${p = \ceil{\frac{1}{2}(1-\epsilon)\log_q(n)}}$ for any $0<\epsilon<1$ yields $p \parenv[\big]{1-\frac{2p}{q^{2p}}}^{\floor{n/2p}} = o(1)$, hence based on Theorem~\ref{th::cq_cover} we arrive at the following theorem.

\begin{theorem}
    The redundancy of any $1$-substitution $(\ell,1;1)$-\wread code is bounded from below by 
    \[
	\log_q\log_q(n) - \log_q\binom{q}{2} - o(1).
    \] \label{th::red}
\end{theorem}

\section{Single Substitution Read Codes} \label{sec::construction}


It is already implied by Corollary~\ref{cor::rec_modq} that a redundancy of $t\log n$ symbols suffices to correct at most $t$ substitutions in the $(\ell,1)$-\wread vector. However, according to Theorem~\ref{th::red}, a more efficient code may exist for the $t=1$ case. This section introduces such a construction that is of optimal redundancy up to an additive constant. 

We define a specific permutation for any $\bfx \in \Sigma_q^n$ as well as its $(\ell,1)$-read vector $\tr[\ell,\delta]{\bfx}$ as
    \begin{align*}
        \bfx^{\pi} &\triangleq C^0_{\ell,1}(\bfx) \circ C^1_{\ell,1}(\bfx) \circ \cdots  \circ C^{\ell-1}_{\ell,1}(\bfx),\\
        \trw{\bfx} &\triangleq\mathcal{R}^{0}(\bfx) \circ \mathcal{R}^{1}(\bfx) \circ \cdots \circ \mathcal{R}^{\ell-1}(\bfx),
    \end{align*}
where ${\mathcal{R}^{i-1}(\bfx)=(\tri{\bfx}{i}, \tri{\bfx}{i+\ell}, \ldots, \tri{\bfx}{i+k\ell})}$ where $k=\floor{\frac{n+\ell-1-i}{\ell}}$ for all $i \in [\ell]$. Recall from Definition~\ref{def::cal} that $C_{\ell,1}^{\alpha}(\bfx)$ refers to a subsequence of $\bfx$.


\begin{env_example}
    Reconsidering $\bfx=(1,2,0,1,2,2)$, we may verify from Examples~\ref{eg::read} and \ref{eg::subsequences} that
    \begin{align*}
        \bfx^{\pi}&=(1,1,2,2,0,2), \\
        \trw{\bfx} &= (1,012,2^2,12,012,2,012,12^2).
    \end{align*}

\end{env_example}

To simplify presentation, we also define the following.

\begin{defn} \label{def::rll}
	Let $a$-$\RLL_q(n)$ 
    be the set of all length-$n$ $q$-ary vectors 
    whose runs are of length at most~$a$. 
\end{defn}

\begin{defn}
    For $n,a>0$, let $\mathcal{H}_q(n,a)$ be the $q$-ary linear code of length $n$, defined by the parity-check matrix
	\begin{equation*}
		\underbrace{\begin{bmatrix}
		\bfH_a & \bfH_a &\cdots & \bfH_a 
		\end{bmatrix}}_{\frac{n(q-1)}{q^a-1} \text{ times}},
	\end{equation*}
	where $\bfH_a$ 
    represents the parity-check matrix of a Hamming code of order $a$, i.e., $\bfH_a$ forms a projective representative (up to a scalar multiple) of all non-zero vectors in $\Sigma_q^a$. 
\end{defn}

Finally, we propose the following code to correct a single substitution in $(\ell,1)$-\wread vectors for $\ell \geq 3$. 
\begin{constr}
    \begin{align*}
        \mathcal{C}(n,\ell)=&\>\{\bfx \in \Sigma_q^n : C^{i}_{\ell,1}(\bfx) \in (\log_q qn)\text{-}\RLL_q(k_i) \; \forall \; i\in \Sigma_\ell, \\
        & \hspace{-14mm}|\trw{\bfx}|_1 \! \bmod q \! \in \!\mathcal{H}_q \big(n+
        \ell \! -\! 1,\log_q\! ( 2(q-1)\log_q (q^2n)\! +\! 1 \big) \},
    \end{align*}  
    \text{where} $k_i=\floor{\frac{n-i-1}{\ell}}+2$.
    \label{con::1sub}
\end{constr}

\emph{Remark:} Note that for $q=2$, the above construction is similar to that defined in Construction~1 of the conference version of this work.


\begin{lemma}
    The redundancy of $\mathcal{C}(n,\ell)$ is at most $$\log_q\log_q n + \log_q\Big( 2(q-1)+\frac{4q-3}{\log_q n}\Big)+1.$$
\end{lemma}

\begin{IEEEproof}
    Observe that the $a\text{-}\RLL_q(n)$ constraint as specified in Definition~\ref{def::rll} is equivalent to the $(d,k)\text{-}\RLL$ constraint \cite{marcusIntroductionCodingConstrained2001}, i.e., restricting each zero run to be of length at least $d=0$ and at most $k=a-1$. Now since the first constraint in Construction~\ref{con::1sub} implies that $C^0_{\ell,1}(\bfx) \circ \cdots \circ C^{\ell-1}_{\ell,1}(\bfx)$ belongs to a superset of $(\log qn)\text{-}\RLL(n)$, we deduce that this run-length restriction necessitates a redundancy of under one symbol, as indicated by \cite[Section~III-B]{levyMutuallyUncorrelatedCodes2019}.

    Next, the second constraint in Construction~\ref{con::1sub}, designed to correct a single substitution error in any contiguous window of length $2\log (qn)+2$ symbols in $|\trw{\bfx}|_1 \bmod q$, corresponds to a parity-check matrix composed of $\frac{(n+\ell-1)(q-1)}{q^a-1}$ copies of $\bfH_a$, which has $2\log (qn)+2=2\log (q^2n)$ columns. Hence,
    \begin{IEEEeqnarray*}{+rCl+x*}
        \frac{q^a-1}{q-1}
        &=& 2\log (q^2n)\\
        \implies a &=& \log \Big( 2(q-1)\log (q^2n)+1 \Big) \\
        &=& \log \Big( 2(q-1) \log n + 4(q-1) +1 \Big) \\
        &=&  \log \log n + \log \Big(2(q-1)+\frac{4q-3}{\log n} \Big).
    \end{IEEEeqnarray*}    
    
    Finally, using the pigeonhole principle, we conclude that $\mathcal{C}(n,\ell)$, which lies in the intersection of the two codebooks corresponding to each of the aforementioned constraints respectively, requires at most $a+1$ redundant symbols, thus proving the statement of the lemma. 
\end{IEEEproof}

To prove that $\mathcal{C}(n,\ell)$ is a $1$-substitution $(\ell,1;1)$-\wread code, we first show how the inherent characteristics of read vectors reveal some information on the substitution error.




\begin{lemma}
    If a substitution error affects the $(\ell,1)$-\wread vector of some $\bfx \in \Sigma_q^n$ where $\ell\geq 3$, thus producing a noisy copy $\tr{\bfx}'$, then there exist $\alpha, \beta \in \Sigma_\ell$ where $\alpha \equiv (\beta+1) \bmod \ell$, such that $\prod_{i}\dtr{\bfx}{\beta}'_i= \big(\prod_{i} \dtr{\bfx}{\alpha}'_i\big)^{-1} \neq c(\cnull)$, and for all $\gamma \not\in\{\alpha, \beta\}$, $\prod_i \dtr{\bfx}{\gamma}'_i=c(\cnull)$. This implies that
    \begin{enumerate}
        \item the composition error is 
        \begin{align*}
            \prod_{i} \big( \tri{x}{i}'\cdot \tri{x}{i}^{-1} \big) &=\prod_{i}\dtr{\bfx}{\beta}'_i= \big(\prod_{i} \dtr{\bfx}{\alpha}'_i\big)^{-1};
        \end{align*}
    	\item the error occurred at an index ${k \in [n+\ell-1]}$, where $k \mathrel{\equiv} \alpha \pmod{\ell}$.
    \end{enumerate}
    \label{lem::sub_err_val_ind}
\end{lemma}

\begin{IEEEproof}
	Suppose the concerned substitution error occurs at index $k \in [n+\ell-1]$. Thus, the noisy \wread vector can be expressed as
	${\tr{\bfx}'=(\tri{x}{1}', \ldots, \tri{x}{n-\ell+1}')}$, where $\tri{x}{k}'\neq \tri{x}{k}$ and $\tri{x}{p}'=\tri{x}{p}$ for all $p \neq k$. 
	
	Denoting $\alpha \triangleq k\bmod \ell$, $\beta \triangleq (k-1)\bmod \ell$, 
    observe that $\dtr{\bfx}{\beta}'$ and $\dtr{\bfx}{\alpha}'$ no longer uphold Lemma~\ref{lem::sum_d0}. Instead,  
    \begin{align*}
        \prod_{i} \dtr{\bfx}{\beta}'_i&=\prod_{i} \dtr{\bfx}{\alpha}^{'-1}_i \\
        &=\tri{x}{k}' \cdot \tri{x}{k}^{-1},
    \end{align*}
	which is the composition error. The preceding equation suggests that the error occurred somewhere in $\mathcal{R}^{\beta}(\bfx)'$, which is a subsequence of $\tr{\bfx}'$. 
    Alternatively, we say that the decoder can only infer the the error position up to the modulo class of~$k$.
\end{IEEEproof}

Next, we show that some composition substitutions are trivial to correct.

\begin{lemma}
    Say a composition substitution corrupts the $i$-th index of $\tr{\bfx}$ to $\tri{x}{i}'$. This error is readily correctable if any of the following conditions holds 
    \begin{enumerate}
        \item Denoting $\tri{x}{i}'
        \mathrel{=} 0^{i_0}\cdots (q-1)^{i_{q-1}}$, 
        if it does not hold that
        $0\leq i_j \leq \ell$ for all $j \in \Sigma_q$, and $\sum_{j=0}^{q-1}i_j=\min \lbrace i,\ell, n-i+1 \rbrace$.
        \item At least one of $\tri{x}{i}'\tri{x}{i-1}^{-1}$ or $\tri{x}{i+1}\tri{x}{i}^{'-1}$ is neither $\cnull$ nor of the form $a\cdot b^{-1}$ for any $a,b \in \Sigma_q$. 
    \end{enumerate} \label{lem::err_geq_2p}
\end{lemma}

\begin{IEEEproof}
    Suppose the error occurred at index $k$. Then, we may express the noisy \wread vector as ${\tr{\bfx}'=(\tri{x}{1}', \ldots, \tri{x}{n+\ell-1}')}$, where $\tri{x}{k}'\neq \tri{x}{k}$ and $\tri{x}{p}'=\tri{x}{p}$ for all $p \neq k$. 
    
    Since in the first case it follows directly from Definition~\ref{def::rv} that the error can be detected and corrected by Corollary~\ref{cor::rec_modq}, we direct our attention to the second case. On account of $\delta=1$, we know that for any $p\in [n+\ell-2]$, it should hold that $\tri{x}{p+1}\cdot \tri{x}{p}^{-1}=x_{p+1} x_{p+1-\ell}^{-1}$, which is either evaluates to $\cnull$ or stays in the form $a\cdot b^{-1}$, where $a,b \in \Sigma_q$ and $a\neq b$. Say $\tri{x}{i}'\tri{x}{i-1}^{-1}$ violates this. As a result, we immediately infer that $k \in \{i-1,i\}$. However, since $\tri{\bfx}{i}' \cdot \tri{\bfx}{i-1}^{-1}$ and $i \bmod \ell$ can be deduced due to Lemma~\ref{lem::sub_err_val_ind}, we are able to conclude that $k=i$, and thereby correct the error.
\end{IEEEproof}


\begin{env_example}

$\tr[3,1]{v}'=(1,12,012,012,2^3,12^2,2^2,2)$ arises from a single substitution in the $(3,1)$-read vector of some $\bfv\in\Sigma_3^6$. Since $\tri{v}{5}'\tri{v}{4}^{'-1}=0^{-1}2^21^{-1}$, we know that either $\tri{v}{4}'$ or $\tri{v}{5}'$ is erroneous. Also, since $\prod_i \dtr{v}{1}'_i=\big( \prod_i \dtr{v}{2}'_i\big)^{-1}=0^{-1}1^{-1}2^2$, we use Lemma~\ref{lem::sub_err_val_ind} to conclude that the composition error is $0^{-1}1^{-1}2^2$ and that the error location, say $k$, satisfies $k \bmod \ell=2$. Thus, we can reverse the substitution error by applying $\tri{v}{5} \leftarrow \tri{v}{5}'\cdot (0^{-1}1^{-1}2^2)^{-1}$, to finally obtain $\tr[3,1]{v}=(1,12,012,012,012,12^2,2^2,2)$, which corresponds to $\bfv=(1,2,0,1,2,2)$.

\end{env_example}

Due to Lemma~\ref{lem::err_geq_2p}, we focus for the rest of the section on proving that $\mathcal{C}(n,\ell)$ can correct a single substitution that is not readily correctable by Lemma~\ref{lem::err_geq_2p}. Next, we demonstrate that the index of such substitutions may be narrowed down.

\begin{env_example}
	$\tr[3,1]{\bfv}'=(1,12,012,02^2,012,12^2,2^2,2)$ arises from a substitution in the $(3,1)$-\wread vector of some $\bfv \in \Sigma_3^{6}$. 
    As $\prod_i \dtr{\bfv}{0}'_i=\prod_i \dtr{\bfv}{1}^{'-1}_i=1^{-1}2$, Lemma~\ref{lem::sub_err_val_ind} suggests that the erroneous composition differs from the true composition by a factor of $1^{-1}2$ and occurred somewhere in $(\tri{v}{1}', \tri{v}{4}', \tri{v}{7}')$. Now assigning ${\tri{v}{1}' \leftarrow \tri{v}{1}'\cdot 12^{-1}}$ yields an invalid read vector since by definition, ${\tri{\bfx}{1} \in \Sigma_q}$. On the contrary, assigning ${\tri{v}{4}' \leftarrow \tri{v}{4}' \cdot 12^{-1}}$ or ${\tri{v}{7}' \leftarrow \tri{v}{7}' \cdot 12^{-1}}$ alters $\tr{\bfv}'$ into the $(3,1)$-\wread vector of ${\bfv=(1,2,0,1,2,2)}$ or ${\bfv=(1,2,0,2,1,2)}$ respectively. \label{eg::err_1}
\end{env_example}

Henceforth, we represent the subsequence reconstructed using Corollary~\ref{prop::dtr_sum_first_m} from left to right with a noisy read sub-derivative, say $\dtr{\bfx}{\beta}'$, as ${\widehat{\bfx}^{(\beta)}\triangleq(\widehat{x}_{\beta+1}, \widehat{x}_{\beta+1+\ell}, \ldots, \widehat{x}_{\beta+1+\floor{\frac{n-\beta-1}{\ell}}\ell})}$. Analogously, $\widetilde{\bfx}^{(\beta)}$ corresponds to right to left reconstruction.

\begin{lemma}
    Let a substitution at index $k$ on $(\ell,1)$-\wread vector of $\bfx \in \Sigma_q^n$ where $\ell \geq 3$, produce $\tr{\bfx}'$. For $\beta\triangleq {(k-1) \bmod \ell}$, if there exists $i>0$ such that $\tri{\bfx}{k+i\ell}'\neq\tri{\bfx}{k+i\ell-1}'$, then $\bracenv*{\widehat{x}_k, \widehat{x}_{k+i\ell}}\not\subseteq \Sigma_q$, where $\widehat{x}_k, \widehat{x}_{k+i\ell}$ are elements of ${\widehat{\bfx}^{(\beta)}}$. 
    \label{lem::ltr_rec}
\end{lemma}
\begin{IEEEproof}
    Since $\tri{x}{k}'$ alone is erroneous, we infer that $(\widehat{x}_{\beta+1}, \widehat{x}_{\beta+\ell+1}, \ldots, \widehat{x}_{k-\ell})=(x_{\beta+1}, x_{\beta+\ell+1}, \ldots, x_{k-\ell})$ and $\widehat{x}_{k+i\ell} \cdot x_{k+i\ell}^{-1} =\tri{x}{k}'\cdot \tri{x}{k}^{-1}$ for all $i\geq 0$. In the following, let $e\triangleq\tri{x}{k}'\cdot \tri{x}{k}^{-1}$. Recall that we consider only such substitution errors that are not readily correctable by Lemma~\ref{lem::err_geq_2p}. 

    Assume $\widehat{x}_k\in \Sigma_q$ and $i > 0$ is minimal such that $\tri{\bfx}{k+i\ell}'\neq\tri{\bfx}{k+i\ell-1}'$; 
    Note that, equivalently, $\tri{\bfx}{k+i\ell}\neq\tri{\bfx}{k+i\ell-1}$
    ; i.e., for all $0<j<i$, 
    \begin{align*}
        \dtr[\ell,\delta]{\bfx}{\beta}'_{\frac{k-\beta-1}{\ell}+j+1} 
        = \dtr[\ell,\delta]{\bfx}{\beta}_{\frac{k-\beta-1}{\ell}+j+1} 
        = c(\cnull)
    \end{align*}
    and $\dtr[\ell,\delta]{\bfx}{\beta}'_{\frac{k-\beta-1}{\ell}+i+1} = \dtr[\ell,\delta]{\bfx}{\beta}_{\frac{k-\beta-1}{\ell}+i+1} = x_{k+i\ell} x_k^{-1}$, where $x_{k+i\ell}\neq x_k$. It follows that 
    \begin{align*}
        \widehat{x}_{k+i\ell} 
        &= \prod_{j=1}^{\frac{k-\beta-1}{\ell}+i+1} \dtr[\ell,\delta]{\bfx}{\beta}'_j \\
        &= \widetilde{x}_k\cdot \prod_{j=1}^i \dtr[\ell,\delta]{\bfx}{\beta}'_{\frac{k-\beta-1}{\ell}+j+1} \\
        &= \widetilde{x}_k\cdot (x_{k+i\ell} x_k^{-1}), 
    \end{align*}
    and since by assumption $\widetilde{x}_k \in \Sigma_q\setminus \bracenv*{x_k}$ and $x_k\neq x_{k+i\ell}$, we have $\widehat{x}_{k+i\ell}\not\in \Sigma_q$.
\end{IEEEproof}

\begin{corollary}
Let $\tr{\bfx}'$ arise from a substitution at index $k$ on $(\ell,1)$-\wread vector of $\bfx \in \Sigma_q^n$ where $\ell \geq 3$. For $\beta\triangleq {(k-1) \bmod \ell}$, if there exists $j
\mathrel{>}
0$ such that $\tri{\bfx}{k
\mathbin{-}
j\ell}'\neq\tri{\bfx}{k
\mathbin{-}
j\ell-1}'$, then $\bracenv*{\widetilde{x}_k, \widetilde{x}_{k
\mathbin{-}
j\ell}}\not\subseteq \Sigma_q$, where $\widetilde{x}_k, \widetilde{x}_{k
\mathbin{-}
j\ell}$ are elements of ${\widetilde{\bfx}^{(\beta)}}$. \label{cor::rtl_rec}
\end{corollary}

A consequence of Lemma~\ref{lem::sub_err_val_ind} and the preceding results is that reconstruction with any corrupted \wread subderivative from left to right and right to left, might help us narrow in on the position of the substitution error. This is stated more formally as follows. 

\begin{lemma}
	For $\ell \geq 3$, let $\tr{\bfx}'$ be a noisy $(\ell,1)$-\wread vector of $\bfx \in \Sigma_q^n$, such that for some $\alpha,\beta \in \Sigma_\ell$, where ${\alpha\equiv \beta
    \mathbin{+}1 \pmod \ell}$, 
    $\prod_i\dtr{\bfx}{\beta}'_i=\prod_i \dtr{\bfx}{\alpha}^{'-1}_i \neq c(\cnull)$. Reconstruction by Corollary~\ref{prop::dtr_sum_first_m} with $\dtr{\bfx}{\beta}'$ from left to right (respectively, right to left) yields $\widehat{\bfx}^{(\beta)}$ ($\widetilde{\bfx}^{(\beta)}$) for which we define $i$ ($j$) as the minimum (maximum) index at which ${\widehat{x}_{\beta+i\ell+1}\not\in \Sigma_q}$ ($\widetilde{x}_{\beta+j\ell+1}\not\in \Sigma_q$), or $i=\floor{\frac{n-\beta-1}{\ell}}+1$ ($j=-1$) if no such index exists. 
    Then, it holds that for all $j+ 1 \mathrel{<} h \mathrel{<} i$, $\tri{\bfx}{\beta+h\ell+1}'=\tri{\bfx}{\beta+h\ell}'$ and the error position in $\tr{\bfx}'$, say $k$, satisfies $\frac{k-\beta-1}{\ell} \in \{j+1, j+2, \ldots, i\}$. \label{lem::recon_halt}
\end{lemma}

\begin{env_example}
	We reconsider $\tr[3,1]{\bfv}'$ from Example~\ref{eg::err_1}. From $\dtr{\bfv}{0}' = (1,1^{-1}2,1^{-1})$, we reconstruct $\widehat{\bfv}^{(0)}=(1,2)$ and $\widetilde{\bfv}^{(0)}=(1^22^{-1},1)$. Since $\widehat{\bfv}^{(0)} \in \Sigma_3^2$ and $\widetilde{v}_1 \not\in \Sigma_3$, we set $i=2$ and $j=0$ in accordance with Lemma~\ref{lem::recon_halt}. Thus, either $\tri{x}{4}'$ or $\tri{v}{7}'$ is noisy, implying that $\bfv=(1,2,0,1,2,2)$ or $\bfv=(1,2,0,2,1,2)$ respectively.
\end{env_example}


Lemma~\ref{lem::recon_halt} essentially suggests that attempting reconstruction with a noisy read sub-derivative may help to narrow down the error location even further. This finally allows us to arrive at

\begin{theorem}
	For $\ell \geq 3$, $\mathcal{C}(n,\ell)$ is a $1$-substitution $(\ell,1;1)$-\wread code.
\end{theorem}
\begin{IEEEproof}
	Let $\tr{\bfx}'$ arise from a single substitution on $(\ell, 1)$-\wread vector of some $\bfx \in \mathcal{C}(n,\ell)$. In light of Lemma~\ref{lem::err_geq_2p}, 
    this proof is dedicated to composition errors of the form $ab^{-1}$.
	
	Upon identifying $\alpha,\beta \in \Sigma_\ell$ where ${\alpha\equiv\beta \mathbin{+}1 \pmod \ell}$, 
    such that ${\prod_i\dtr{\bfx}{\beta}'_i=\prod_i \dtr{\bfx}{\alpha}^{'-1}_i \neq c(\cnull)}$, we attempt reconstruction with $\dtr{\bfx}{\beta}'$ from left to right and from right to left to obtain $\widehat{\bfx}^{(\beta)}$ and $\widetilde{\bfx}^{(\beta)}$ respectively, and define indices $i$ and $j$ according to Lemma~\ref{lem::recon_halt}. Since for all ${j+1 < h < i}$, $\tri{\bfx}{\beta+h\ell+1}'\cdot\tri{\bfx}{\beta+h\ell}^{'-1}=\cnull$, and a run of `$\cnull$'s in $\dtr{x}{\beta}'$ can be of length at most $2\log_q (qn) \mathbin{-} 1$ as a consequence of  the run-length constraint in $\mathcal{C}(n,\ell)$ and Lemma~\ref{lem::cal-bij},
    we infer that $i-j-2 \leq {2\log_q (qn)
    \mathbin{-} 1}$. 
 
    From Lemma~\ref{lem::recon_halt}, we know that the error exists somewhere in $(\tri{\bfx}{\beta+(j+1)\ell+1}', \tri{\bfx}{\beta+(j+2)\ell+1}', \ldots, \tri{\bfx}{\beta+i\ell+1}')$, which is evidently a substring of $\trw{\bfx}'$ and has a length of at most ${2\log_q (qn)} 
    \mathbin{+} 1$. Since an error of the form $ab^{-1}$, where $a\neq b$, surely reflects as a single substitution in $|\trw{\bfx}'|_1 \bmod q$, which belongs to a code that corrects a substitution error localized to a window of ${2\log (qn)} +2$ symbols, 
    we can uniquely recover $|\trw{\bfx}|_1 \bmod q$, and by Corollary~\ref{cor::rec_modq}, also~$\bfx$.
\end{IEEEproof}






\section{Error correction with multiple reads} \label{sec::two_reads}


We first consider the following lemma to see if and how multiple noisy reads might be leveraged to construct more efficient codes for correcting errors in $(\ell, \delta)$-read vectors.

\begin{lemma}
    Exactly one of the following conditions holds for $\ell \geq 3$ and any two distinct $\bfx, \bfy \in \Sigma_q^n$.
    \begin{enumerate}
        \item $d_H(\tr{\bfx}, \tr{\bfy})=2$ and $|B_1(\tr{\bfx}) \cap B_1(\tr{\bfy})|=2$; or
        \item $d_H(\tr{\bfx}, \tr{\bfy})>2$ and $|B_1(\tr{\bfx}) \cap B_1(\tr{\bfy})|=\emptyset$.
    \end{enumerate} \label{lem::ball_overlap}
\end{lemma}

\begin{IEEEproof}
    Since Theorem~\ref{th::no_dh1} already precludes the possibility of $d_H(\tr{\bfx}, \tr{\bfy})=1$ and the case of ${d_H(\tr{\bfx}, \tr{\bfy})>2}$ follows from the triangle inequality, 
    we proceed to prove the remaining case wherein $d_H(\tr{\bfx}, \tr{\bfy})=2$, i.e., $\bfx, \bfy$ satisfy the conditions stated in Theorem~\ref{th::xy1001}.

    More specifically, there exist distinct $i,j \in [n+\ell-1]$ such that $\tri{\bfx}{i}\cdot\tri{\bfy}{i}^{-1}=\tri{\bfy}{j}\cdot\tri{\bfx}{j}^{-1}\neq \cnull$ and for all $k \not\in \{i,j\}$, $\tri{\bfx}{k}=\tri{\bfy}{k}$. This implies that $B_1(\tr{\bfx}) \cap B_1(\tr{\bfy})$ 
    is exactly the following.
    \begin{align*}
    &\{(\tri{\bfx}{1}, \ldots, \tri{\bfx}{i-1}, \tri{\bfy}{i}, \tri{\bfx}{i+1}, \ldots, \tri{\bfx}{n+\ell-1}), \\
                           &(\tri{\bfx}{1}, \ldots, \tri{\bfx}{j-1}, \tri{\bfy}{j}, \tri{\bfx}{j+1}, \ldots, \tri{\bfx}{n+\ell-1})    \} \\
                           &= \{(\tri{\bfy}{1}, \ldots, \tri{\bfy}{j-1}, \tri{\bfx}{j}, \tri{\bfy}{j+1}, \ldots, \tri{\bfy}{n+\ell-1}), \\
                           &(\tri{\bfy}{1}, \ldots, \tri{\bfy}{i-1}, \tri{\bfx}{i}, \tri{\bfy}{i+1}, \ldots, \tri{\bfy}{n+\ell-1})    \}. 
    \end{align*} Hence, the first case directly follows.
\end{IEEEproof}

As for the case of $t=1$, $M=1$ in Section~\ref{sec::lb_red}, we wish to derive a lower bound on the redundancy required by a $t$-substitution $(\ell,1;2)$-read code by constructing a clique cover over a graph $G(n)$ that contains $q^n$ vertices, each corresponding to a specific vector in $\Sigma_q^n$. Two vertices representing the binary vectors $\bfx, \bfy \in \Sigma_q^n$ in $G(n)$ are considered to be adjacent if and only if $|B_1(\tr{\bfx}) \cap B_1(\tr{\bfy})|\geq 2$.

Since Lemma~\ref{lem::ball_overlap} suggests that $G(n)$ is identical to $\mathcal{G}(n)$ as defined in Section~\ref{subsec::cq}, we infer that for any positive integer $p$, the clique-cover $\mathcal{Q}_p$ for $\mathcal{G}(n)$ from Theorem~\ref{th::cqc_specified}, also acts as a clique-cover for $G(n)$. As a consequence, we arrive at the following lemma.

\begin{lemma}
    The redundancy of a $1$-substitution $(\ell,1;2)$-\wread code is bounded from below by 
    \[
	\log_q\log_q(n) \mathbin{-} \log_q\binom{q}{2} - o(1).
    \] \label{th::red_2}
\end{lemma}

Thus, $\mathcal{C}(n,\ell)$ is also a $1$-substitution $(\ell, 1; 2)$-\wread code that is of optimal redundancy up to an additive constant. 

Evidently, given three distinct noisy copies of the $(\ell, 1)$-\wread vector of any $\bfx \in \Sigma_q^n$, one can uniquely reconstruct $\tr{\bfx}$ and thereby $\bfx$.

\section{Conclusion} \label{sec::concl}

The primary objective of this work was to initiate a line of research dedicated to error-correcting codes that attempt to incorporate the dominant physical aspects of nanopore sequencing. The channel model we adopted incorporates the intersymbol interference aspect of the sequencer as a window that slides over the incoming DNA strand and outputs the composition of the corresponding substrings in this strand. The measurement noise in the current readout is modeled as substitution errors in the resulting vector of compositions. We observed how, in doing so, the correction of a single substitution can be accomplished with $\log_q\log_q n \mathbin{+}O(1)$ 
redundant symbols, instead of $\log_q n$ symbols necessitated by the standard case, i.e., when the decoder is agnostic to the channel model. This result understandably encourages us to further investigate this channel model under multiple substitution errors as well as more error settings, e.g., deletions and duplications. Examining this channel in the context of Levenshtein's sequence reconstruction problem is also an exciting avenue to pursue.

\bibliographystyle{IEEEtran}
\bibliography{ISIT/literature.bib}


\appendix

\subsection{Proof of \texorpdfstring{Lemma~\ref{lem::ccq}}{Lemma~7}, non-binary extension of \texorpdfstring{\cite[Lemma 7]{chrisnataCorrecting2022}}{[8, Lemma~10]}}

\begin{IEEEproof}
    Since all singletons are cliques, we endeavor to show that for all $\boldsymbol{\gamma}=(\bfu,\bfw,a,b) \in \Gamma$, $\mathcal{Q}_{\gamma}^{(0)}$ is a clique. The proof for $\mathcal{Q}_{\gamma}^{(1)}$ follows similarly.
    
    For any two vectors in $\mathcal{Q}_{\gamma}^{(0)}$, say $\bfx=\bfu(ab)^i(ba)^{p-i} \bfw$ and $\bfy=\bfu(ab)^j(ba)^{p-j} \bfw$, we may assume $i<j$ without loss of generality, and observe that
    \begin{align*}
        \bfx&=\bfu(ab)^i(ba)^{j-i}(ba)^{p-j} \bfw, \\
        \bfy&=\bfu(ab)^i(ab)^{j-i}(ba)^{p-j} \bfw.
    \end{align*}

    By Definition~\ref{def::gp}, $\bfx$ and $\bfy$ are clearly adjacent, implying that $\mathcal{Q}_{\gamma}^{(0)}$ is a clique. 

    Now to show that each vector $\bfx \in \Sigma_q^{2pm}$ belongs to at least one clique in $\mathcal{Q}(m,p)$, note that we either have $\bfx \in \widetilde{\Lambda}_p^m $, or one of the $m$ subblocks of $\bfx$ lies in $\Lambda_{p,a,b}$, for some $a,b \in \Sigma_q$. In the former case, $\bfx$ constitutes a singleton and is accounted for by $\mathcal{Q}(m,p)$, while in the latter case, assuming that the $i$th subblock is the first that lies in $\Lambda_{p,a,b}$, we deduce that $\bfx$ belongs to the clique $Q_{(\bfu,\bfw,a,b)}^{(0)}$ where $\bfx_1^{2p(i-1)}=\bfu \in \widetilde{\Lambda}_{p,a,b}^{i-1}$, while $\bfx_{2pi+1}^{2pm}=\bfw \in \Sigma_q^{2p(m-i)}$.
\end{IEEEproof}

\subsection{Non-binary extension of \texorpdfstring{\cite[Lemma 9]{chrisnataCorrecting2022}}{[8, Lemma~13]}} \label{app::o1}

\begin{lemma}
    For $p = \ceil{\frac{1}{2}(1-\epsilon)\log_q(n)}$, we have $\lim_{n \to \infty} p\Big(1-\frac{2p}{q^{2p}} \Big)^{\floor{\frac{n}{2p}}}=0$.
\end{lemma}

\begin{IEEEproof}
Based on $1-x\leq e^{-x}$ we observe 
\begin{IEEEeqnarray*}{+rCl+x*}
        p \parenv[\Big]{1-\frac{2p}{q^{2p}}}^{\floorenv*{n/2p}} 
        &\leq& p \exp\parenv*{-\frac{2p}{q^{2p}} \parenv[\Big]{\frac{n}{2p}-1}} \\
        &=& p \exp\parenv*{-\frac{n-2p}{q^{2\ceilenv{\frac{1}{2} (1-\epsilon) \log(n)}}}} \\
        &\leq& p \exp\parenv*{-\frac{n-2p}{q^2 n} n^\epsilon} \\
        &\leq& \log(n) \cdot \exp\parenv*{-\frac{n-\log(n)}{q^2 n} n^\epsilon} \xrightarrow{n\to \infty} 0.
\end{IEEEeqnarray*}
\end{IEEEproof}

\end{document}